\documentclass[12pt]{article}
\usepackage{amsfonts,latexsym,amssymb,graphicx}

\def\d{{\rm d}}
\def\mi{{\rm i}}
\def\G{\mathop{\Gamma}\nolimits}
\def\l{\lambda}
\def\Re{\mathop{\rm Re\,}\nolimits}
\def\e{\mathop{\rm e}\nolimits}
\def\Res{\mathop{\rm Res}\nolimits}
\def\hf{{\textstyle\frac{1}{2}}}
\def\qt{{\textstyle\frac{1}{4}}}
\def\sq2{{\textstyle\frac{1}{\sqrt 2}}}
\def\od{{\textstyle\frac{1}{2}+\frac{1}{N}}}

\def\defi{\stackrel{\rm def}{=}}
\def\si{\!\!\! &}
\def\se{& \!\!\!}
\def\beq{\begin{equation}}
\def\eeq{\end{equation}}
\def\bea{\begin{eqnarray}}
\def\eea{\end{eqnarray}}

\title{From exact-WKB toward\\
singular quantum perturbation theory II}

\author{{\bf Andr\'e Voros}$^1$\\
\\
CEA, Service de Physique Th\'eorique de Saclay\\
CNRS URA 2306\\
F-91191 Gif-sur-Yvette CEDEX, France\\
{ E-mail : {\tt voros@spht.saclay.cea.fr}}\\}

\begin{document}
\maketitle

\setcounter{footnote}{1}
\footnotetext{Also at: Institut de Math\'ematiques de Jussieu--Chevaleret, 
CNRS UMR 7586, Universit\'e Paris 7, F-75251 Paris CEDEX 05, France.}

\begin{abstract}
Following earlier studies, several new features of singular perturbation 
theory for one-dimensional quantum anharmonic oscillators are computed 
by exact WKB analysis; former results are thus validated.
\end{abstract}

This note continues our study \cite{AVK} of singular perturbation theory 
in one-dimensional (1D) quantum mechanics using exact WKB analysis. 
Our focus remains the $v \gg 1$ regime for the potentials $V(q) = q^N + vq^M$ 
on the real line, with $N>M$ positive even integers. 
Among those, the \emph{quartic oscillator} $q^4+vq^2$
has been a prime model for the mathematics 
of quantum perturbation theory \cite{AVBW,AVSi,AVGS,AVHM,AVSh}.
Kawai and Takei \cite{AVKT} pioneered the use of exact WKB analysis 
in the latter context, followed by~\cite{AVDP} (see also
\cite[Introduction to Part~I, and Pham's contribution]{AVKB}, \cite{AVZJ},
and references therein).
In spite of those successes, present exact-WKB \emph{quantization conditions}  
(for ${q^4+vq^2}$, say) \emph{fail to tend} 
toward their harmonic-potential ($vq^2$) counterparts as ${v \to +\infty}$, 
be it analytically or numerically~\cite{AVEX}.
This worrying observation triggered our present line of work 
(starting from \cite[\S~3]{AVQ}): to further probe how consistently 
exact WKB theory handles the perturbative ($v \gg 1$) regime.
\medskip

We are happy and honored to dedicate this work to Professor Kawai with gratitude, 
for his many essential contributions and leadership in exact WKB analysis, 
but also earlier (with Professors Sato and Kashiwara) in hyper/micro/function theory;
this framework greatly inspired, and its Authors warmly encouraged, our first steps 
in exact WKB analysis~\cite{AVQO}.

\medskip

Even though this work is thoroughly tied to \cite{AVK} (with its bibliography),
in \S~\ref{AV1} we recall the main background
and further strengthen the case for \emph{improper} (divergent) action integrals
like $\int_0^\infty \Pi(q) \, \d q$, where $\Pi(q)={(V(q)+\l )^{1/2}}$ 
is the classical momentum function.
In \S~\ref{AV2} we present new cases where $\int_0^\infty \Pi(q) \, \d q$ 
can be computed exactly for some \emph{trinomial} $\Pi(q)^2 \,$:
essentially the \emph{quartic} case $\Pi(q)^2=q^4+vq^2+\l $,
for which $\int_0^\infty \Pi(q) \, \d q$ reduces
to ordinary (i.e., convergent) \emph{complete elliptic} integrals.
In \S~\ref{AV3} we extend the main outcome of \cite{AVK}, 
namely the $v \to +\infty$ asymptotic expression of the spectral determinants
$D_N^\pm(\l ,v) \defi \det^\pm ( -\d^2/\d q^2 + q^N + vq^M + \l )$
in terms of $D_M^\pm(\Lambda) \defi \det^\pm ( -\d^2/\d q^2 + q^M + \Lambda )$, 
to $v \to \infty$ in a \emph{complex sector}.
Thanks to this, finally in \S~\ref{AV4} we demonstrate (currently provided $2M+2>N$)
how the fundamental bilinear functional relation satisfied by $D_N^\pm$
\emph{does evolve} into its counterpart for $D_M^\pm$ as $v \to +\infty$,
in spite of a discontinuous jump at $v=\infty$ of the main parameter, 
the \emph{degree} of the potential (from $N$ to $M$).

\section{Background (summarized) \cite{AVK}}
\label{AV1}

Our model of quantum perturbation theory is the 1D Schr\"odinger equation
\beq
\label{AVSE}
\Bigl[ -\frac{\d^2}{\d q^2} + V(q) + \l \Bigr] \Psi(q)=0, \quad 
q \in {\mathbb R}, \quad V(q) = q^N + v q^M , \quad v \gg 1 ,
\eeq
with $N>M$ positive even integers. 
If we use the unitary equivalence (called \emph{Symanzik scaling})
\beq
\label{AVSS}
-\frac{\d^2}{\d q^2} + uq^N + vq^M \approx v^{2/(M+2)} 
\Bigl[ -\frac{\d^2}{\d x^2} + x^M + u v^{-(N+2)/(M+2)} \, x^N \Bigr] 
\eeq
twice, at $u=1$ and $u=0$, the resulting right-hand sides imply 
that the operator $\hat H=-\d^2/\d q^2+q^N+v q^M$ is a singular perturbation
of $\hat H_0=-\d^2/\d q^2+vq^M$ for $v \gg 1$,
and that the degree \emph{drops} from $N$ to $M$ \emph{at} $v=+\infty$.

From the classical dynamics we will use the \emph{momentum} function 
($\times \, \mi$),
\beq
\Pi (q) = (V(q)+\l )^{1/2} \qquad \mbox{(real in the classically forbidden region)},
\eeq
and its \emph{residue} $\Res_{q=\infty} \Pi (q) = \beta_{-1}(0)$, 
a notation based on the expansion~\cite{AVQ}
\beq
\label{AVBE}
(V(q) + \lambda)^{-s+1/2} \sim
\sum_\rho \beta_\rho (s) \, q^{\rho-Ns} \quad \textstyle
(q \to \infty); \quad \rho= \frac{1}{2} N,\ \frac{1}{2} N -1, \ldots
\eeq

The spectrum of $\hat H$ is purely discrete,
$0<\l _0 < \l _1 < \cdots \uparrow +\infty$,
and separates according to parity since $V$ is an even function.
Useful spectral functions (labeled by parity) are 
the \emph{generalized zeta functions},
\beq
\label{AVZF}
Z^\pm (s,\l ) \defi \sum_{k \ {\rm even \atop odd}} (\lambda_k + \lambda)^{-s} 
\qquad (\Re s > \od) ,
\eeq
and the \emph{spectral determinants} $D^\pm(\l )$, 
defined through \emph{zeta regularization},
\beq
\label{AVSD}
\log D^\pm(\l ) \equiv \log {\det}^\pm (\hat H+\l ) \defi 
[-\partial_s Z^\pm (s,\lambda)]_{s \leadsto 0} ,
\eeq
where ``${s \leadsto 0}$" implies analytical continuation in $s$. Scaling laws follow:
\beq
\label{AVSL}
\begin{array}{rcl}
{\det}^\pm [r (\hat H+\l )] \si \equiv\se 
r^{Z^\pm(s=0,\l )} \, {\det}^\pm (\hat H+\l ) \qquad (\forall r>0) , \\
\mbox{where} \qquad Z^\pm(s=0,\l ) \si \equiv\se 
\displaystyle -\frac{\beta_{-1}(0)}{N} \pm \frac{1}{4} 
\end{array}
\eeq
\cite[equations~(7), (30)]{AVEX}\cite[equations~(15), (27), (37)]{AVQ}.

A more concrete realization of $\log D^\pm$ through (\ref{AVSD}) is, 
first to formally apply $(\d / \d \l )^m$ to (\ref{AVSD}) with the minimal $m$ 
such that the result ($\propto Z(m,\l )$) converges, 
i.e., $m > \frac{1}{2}+\frac{1}{N}$, then to integrate back:
the separate knowledge that the ${\l \to +\infty}$ expansion of $\log D^\pm(\l )$
shall only have ``canonical" terms \cite{AVC}\cite[\S~1.1.2]{AVQ} 
fixes the $m$ integration constants. Here, $N \ge 4$ implies $m=1:$ specifically,
\beq
\frac{\d}{\d \l } \log D^\pm(\l ) \equiv Z^\pm (1,\l )
\eeq
converges according to (\ref{AVZF}), 
and $\log D^\pm(\l )$ is then \emph{the} unique primitive of $Z^\pm (1,\l )$ 
which is devoid of a constant ($\propto \l ^0$) term in its large-$\l $ expansion.
\medskip

\emph{Classical} analogs of those quantum determinants can be defined as well 
\cite{AVEX,AVQ,AVZJ}, through: $\log D_{\rm cl}^\pm(\l ) \defi$ 
\{the divergent part of $\log D^\pm(\l )$ for $\l \to +\infty $\},
or equivalently \cite[\S~1.2.1 and equation~(46)]{AVQ} through:
\bea
\log (D_{\rm cl}^+ / D_{\rm cl}^-) \, (\l ) \si =\se \log \Pi(0)
\equiv \hf \log \l , \\
\label{AVDC}
\log (D_{\rm cl}^+ D_{\rm cl}^-) (\l ) \si =\se 
\int_{-\infty}^{+\infty} \! \Pi(q) \, \d q 
= 2 \, I, \quad I \defi \int_0^{+\infty} \! (V(q) + \l )^{1/2} \, \d q , \quad
\eea
where this divergent ``improper action integral" gets specified 
\emph{just like} $\log D^\pm$: first,
\beq
\label{AVDD}
\frac{\d I}{\d \l }= \frac{1}{2} \int_0^{+\infty} \! (V(q)+\lambda)^{-1/2} \, \d q
\equiv \frac{1}{4} \int_{-\infty}^{+\infty} \! (V(q)+\lambda)^{-1/2} \, \d q
\eeq
converges, then $I(\l )$ is \emph{defined} as that primitive of (\ref{AVDD}) 
which is devoid of a constant ($\propto \l ^0$) term in its large-$\l $ expansion.

\emph{Improper} actions as in (\ref{AVDC}) (i.e., along \emph{infinite} paths) 
offer \emph{many} benefits for asymptotic and exact WKB analysis. 
WKB solutions of (\ref{AVSE}) can now be defined intrinsically: 
e.g., as $\Psi_{\rm WKB}(q) = \Pi(q)^{-1/2} \exp \int_{-\infty}^q \Pi(q') \, \d q'$,
unlike the traditional forms which awkwardly involve extraneous base points. 
The geometrical analysis no longer requires to set infinite paths apart
as it used~to \cite{AVQO,AVDD}.
Moreover, the algebra itself is simplified; 
e.g., consider the full determinant $D(\l) \defi (D^+ D^-)(\l )$: 
previously, to get the large-$\l $ expansion of $\log D$ 
in the simplest case $V(q)=q^{2M}$ \cite{AVQO},
we had to factor $D = D_{\rm cl} \, a \ $ ($a(\l )$ is the ``Jost function"), 
then expand $\log a$ using
$\log a \equiv \int_{-\infty}^{+\infty} [U-\Pi](q) \, \d q$ where 
$\Psi(q) = U(q)^{-1/2} \exp \int U \, \d q$ parametrizes an exact solution 
of (\ref{AVSE}), and finally obtain $\log D_{\rm cl}$ by other means; 
now that the improper integrals (\ref{AVDC}) are allowed, 
all that condenses into a single identity (valid for general~$V$):
\beq
\log D \equiv \int_{-\infty}^{+\infty} \! U(q) \, \d q .
\eeq

\section{Explicit improper actions for trinomial $\Pi(q)^2$}
\label{AV2}

In \cite{AVK}, we computed the improper action integral 
$I=\int_0^{+\infty} \Pi(q) \, \d q$ 
in closed form for \emph{any binomial} $\Pi(q)^2=uq^N+vq^M$;
then (\S~4.2) we stated that we could no longer do so for a \emph{trinomial} 
of the general (even) form $ \Pi(q)^2=q^N+vq^M+\l $ (with ${N>M>0}$),
for which we just needed the $v \to +\infty$ behavior of~$I$ anyway 
\cite[equation~(4.16)]{AVK}, reproduced as (\ref{AVTA}) below.

It is nevertheless \emph{wrong} to infer from the above that 
\emph{strictly no} exact computations can be done in fully trinomial cases,
and we now present several examples
(still for even $q^N+vq^M+\l $ with positive $v,\ \l $).
After recalling the closed-form results for binomials, we will quote 
another, trivial and degenerate, instance: perfect-square trinomials.
Then, our main new case will be the \emph{quartic} anharmonic oscillator: 
we can reduce its improper action exactly 
to standard (i.e., convergent) action integrals, 
and therefrom to \emph{complete elliptic integrals} \cite{AVBW,AVME,AVHM,AVKV},
as (\ref{AVTQ})--(\ref{AVTI}) below;
we then verify the abovementioned large-$v$ behavior on this case ($N=4$).
Finally, the same approach must work for higher-degree polynomial $V(q)$, 
converting $\int_0^{+\infty} \Pi(q) \, \d q$ exactly 
into convergent \emph{hyperelliptic} integrals (as studied in \cite{AVDD,AVE}); 
but since the latter remain not so explicitly understood, 
we will skip this case ($N>4$) here.

\subsection{Binomial $\Pi(q)^2$ : exact evaluation}
For $\Pi(q)^2 = uq^N+vq^M$ (with $N>M \ge 0$),
$\int_0^\infty \Pi(q) \, \d q$ was exactly computed in \cite[\S~4.1]{AVK}. 
We recall the main formulae for later convenience:
\bea
I \si =\se \int_0^{+\infty} \! (uq^N+vq^M)^{1/2} \, \d q \defi \lim_{s \leadsto 0} I_0(s), \\
I_0(s) \si =\se \int_0^{+\infty} (uq^N+vq^M)^{1/2 \, -s} \, \d q 
\qquad (\Re s > \od) \\
\si \equiv\se \frac{\G ( \frac{M(1 - 2s) + 2}{2(N-M)} )
\G ( -\frac{N(1 - 2s) + 2}{2(N-M)} )}{(N-M) \, \G (s \!-\! 1/2) }
\, u^{-\frac{M(1-2s)+2}{2(N-M)}} v^\frac{N(1-2s)+2}{2(N-M)} ,
\eea
where ``${s \leadsto 0}$" implies analytical continuation in $s$, with the result:
\beq
\label{AVBN}
I = \frac{ \G (j-\hf) \G (-j)}{(N-M) \G (-\hf) } \, u^{-j+1/2} \, v^j \quad
\textstyle \bigl( j \defi \frac{N+2}{2(N-M)} > \hf \bigr) 
\quad \hbox{when finite}; 
\eeq
otherwise, i.e., when $j=1,2,\ldots$, 
a further ``canonical" renormalization yields
\beq
\label{AVBA}
\begin{array}{cc}
\displaystyle I = -\frac{ 2j \, \beta_{-1}(0)}{N+2 } 
\Biggl[ \log v - \sum_{m=1}^j \frac{1}{m} - \frac{2M}{N} 
\Bigl( \log 2 + \hf \log u - \sum_{m=1}^{j-1} \frac{1}{2m \!-\! 1} \Bigr) \Biggr],
\\
\displaystyle \beta_{-1}(0) = 
(-1)^{j-1} \frac{(2j-2)!}{2^{2j-1} (j-1)! \, j!} \, u^{-j+1/2} \, v^j .
\end{array}
\eeq

We repeat from \cite{AVK} the examples we will mostly need:
\bea
\label{AVBH}
\int_0^{+\infty} \! (w q^N + \lambda)^{1/2} \,\d q \si = \se 
- \frac{\textstyle \G (1+\frac{1}{N})\G (-\frac{1}{2}-\frac{1}{N})}{2 \sqrt \pi}
\, w^{-\frac{1}{N}} \lambda^{ \frac{1}{2}+\frac{1}{N} } \quad (N \ne 2) \quad \\*
\label{AVB2}
\si = \se -\qt \, w^{-1/2} \lambda (\log \lambda -1) 
\qquad \qquad \qquad \ \ (N =2) \\
\label{AVB4}
\int_0^{+\infty} \! (q^4 + v q^2)^{1/2} \,\d q 
\si = \se - {\textstyle\frac{1}{3}} \, v^{3/2} ,
\eea
all based on (\ref{AVBN}) except (\ref{AVB2}), which uses (\ref{AVBA}) with $j=1$.

\subsection{Perfect-square trinomials: $\Pi(q)^2 = (q^M + \sqrt \l )^2 \ (M \rm{\ even)}$}
This degenerate case trivially reduces to a binomial formula like (\ref{AVBN}), 
using
\beq
\label{AVTB}
\int_0^{+\infty} \! [(q^M + wq^L)^2]^{1/2 \, -s} \d q = \frac
{\G \bigl( \frac{L(1-2s)+1}{M-L} \bigr) \G \bigl( -\frac{M(1-2s)+1}{M-L} \bigr)}
{(M-L) \G (2s-1) } \, w^\frac{M(1-2s)+1}{M-L }
\eeq
for $M>L \ge 0$ and $w>0$. 
The singular formula (\ref{AVBA}) is never needed in our setting ($\Pi(q)^2$ even):
for $M$ and $L$ even, no pole can appear in the numerator of (\ref{AVTB}) 
at $s=0 \,$; but one appears in the denominator instead, leading to
\beq
\label{AVTT}
I=\int_0^\infty [(q^{N/2} + \sqrt \l )^2]^{1/2} \, \d q \equiv 0 
\qquad (\mbox{for even } N/2>0).
\eeq

\subsection{The general even quartic case: $\Pi(q)^2 = q^4+vq^2+\l $}

At present, we mean to exploit the large toolbox of results readily available
for the \emph{complete elliptic integrals} \cite{AVEB,AVBF,AVGR}: 
specifically here,
\beq
\label{AVES}
K(k) \defi \! \int_0^1 \bigl[ (1-t^2)(1-k^2t^2) \bigr] ^{-1/2} \d t , \quad 
E(k) \defi \! \int_0^1 \Biggl[ \frac{(1-k^2t^2)}{(1-t^2)} \Biggr] ^{1/2} \d t ,
\eeq
as functions of the \emph{modulus} $k \,$; 
the \emph{complementary modulus} is $k' \defi \sqrt{1-k^2}$.

\subsubsection*{Main needed formulae}

- special values: \cite[formulae~13.8(5),(6),(15),(16)]{AVEB}
\beq
\label{AVE0}
K(0) = E(0) = \hf \pi
\eeq
\beq
\label{AVE2}
K(\sq2) = \frac{\G (\qt)^2}{4 \sqrt \pi} , \qquad
E(\sq2) = \frac{1}{2} \Biggl[ K(\sq2) + \frac{\pi}{2K(\sq2)} \Biggr] ;
\eeq

\noindent - derivatives: 
\cite[formulae~710.00, 710.02]{AVBF}\cite[formulae~8.123(2),(4)]{AVGR}
\beq
\label{AVED}
\frac{\d K}{\d k} = \frac{E(k)}{k k'^2} - \frac{K(k)}{k} \qquad 
\Bigl( \frac{\d E}{\d k} = \frac{E(k) -K(k)}{k} \mbox{ is not used here} \Bigr) ;
\eeq

\noindent - expansions for $k \to 1^- \iff k' \to 0^+$ \ \ (implying $E(1)=1$):
\cite[p.~93--94]{AVR} \cite[footnote~11 p.~184]{AVBW}
\cite[formulae~900.05, 900.07]{AVBF}
\beq
\label{AVEA}
\begin{array}{l}
\displaystyle K(k) = \log \frac{4}{k'} 
+ \frac{1}{4} \Bigl[ \log \frac{4}{k'} -1 \Bigr] \, k'^2 + O(k'^4 \log k') 
\\[10pt]
\displaystyle E(k) = 1 
+ \frac{1}{2} \Bigl[ \log \frac{4}{k'} - \frac{1}{2} \Bigr] \, k'^2
+ \frac{3}{16} \Bigl[ \log \frac{4}{k'} - \frac{13}{12} \Bigr] \, k'^4 
+ O(k'^6 \log k') 
\end{array}
\eeq

\noindent - selected transformation formulae: \cite[Table~4 p.~319]{AVEB}
\bea
\label{AVT1}
K(k) \si =\se \frac{1+\dot k'}{2} K(\dot k), \quad 
E(k) = \frac{E(\dot k)+ \dot k' K(\dot k)}{1+\dot k'} \qquad 
\mbox{for } k= \frac{1-\dot k'}{1+\dot k'} \\
\label{AVT2}
K(k) \si =\se \tilde k' \, K(\tilde k) , \qquad \ \ E(k) = \frac{E(\tilde k)}{\tilde k'} \qquad 
\qquad \qquad \ \mbox{for } k= \frac{\mi \, \tilde k}{\tilde k'} \, .
\eea

\subsubsection*{Our closed-form result} 
For non-negative $v$ and $\l $ (as in \cite{AVK}, and mainly for simplicity), we find:
\bea
\label{AVTQ}
I \si \defi\se \int_0^{+\infty} \!\! (q^4+vq^2+\lambda)^{1/2} \, \d q \\
\label{AVTR}
(v \ge 2\sqrt \l ): \si \equiv\se \textstyle \frac{1}{3} (v + 2\sqrt \l )^{1/2} 
\bigl[ 2 \sqrt \l K(k) - v E(k) \bigr] , 
\ \ k = \Bigl( \frac{\textstyle v - 2 \sqrt \l}{\textstyle v + 2 \sqrt \l} \Bigr) 
^{1/2} ; \qquad \\
\label{AVTI}
(v \le 2\sqrt \l ): \si \equiv\se \textstyle \frac{1}{3} \, \l ^{1/4} 
\bigl[ (2\sqrt \l + v) K(\tilde k) -2 v E(\tilde k) \bigr] , \ \quad 
\tilde k = \frac{\textstyle (2\sqrt \l - v)^{1/2}}{\textstyle 2 \, \l ^{1/4} } 
\, .
\eea

\noindent \emph{Derivation.} We first specify $\d I/\d \l $ by means of (\ref{AVDD})
for $V(q)={q^4+vq^2}$. In contrast to (\ref{AVTQ}), here the integrand
$(V(q)+\l )^{-1/2}$ \emph{is integrable} at $q=\infty$ in $\mathbb C$,
allowing to deform the path $(-\infty,+\infty)$
to a \emph{bounded} contour in the complex $q$-plane:
\beq
\label{AVTJ}
\frac{\d I}{\d \l } 
= \frac{1}{4} \int_{-\infty}^{+\infty} \! (q^4+vq^2+\lambda)^{-1/2} \, \d q
= \frac{1}{4} \int_C (q^4+vq^2+\lambda)^{-1/2} \, \d q
\eeq
where $C$ is, e.g., a positive contour encircling the pair of roots $\mi q_\pm$ 
of $\Pi(q)^2$ (\emph{turning points}) that lie in the upper half-plane.
We now prefer to pursue explicitly with $v \ge 2\sqrt \l $
(and analytically continue the result to $v \le 2\sqrt \l $ later):
then $0 < q_- \le q_+$, cf.~Fig.~1(a).
The last integral in (\ref{AVTJ}), being taken over a bounded path,
admits a closed-form primitive with respect to $\l $, as
\bea 
\label{AVTC}
\hat I(\l ) \si =\se \frac{1}{2} \int_C (q^4+vq^2+\lambda)^{1/2} \, \d q  \nonumber\\
\label{AVTP}
\si =\se -\int_{q_-}^{q_+} \! (-q^4+vq^2-\lambda)^{1/2} \, \d q
= -{\textstyle\frac{1}{3}} \, q_+ 
\bigl[ vE(\dot k) - 2q_-^{\, 2} K(\dot k) \bigr] \\
\mbox{where} \ \ q_\pm \si =\se 
\bigl[ \hf ( v \pm \sqrt{v^2-4\l }) \bigr]^{1/2} \ \mbox{ and } 
\ \dot k = [ 1 - q_-^{\, 2} / q_+^{\, 2} ]^{1/2}, \ \dot k' = q_-/q_+ \nonumber
\eea
\cite[formula~3.155(1) for $u=b$ and (amplitude) $\l =\pi/2$]{AVEB}
\cite[formula (4.22)\footnote{We think there should be no factor $\rho^{1/2}$ 
on the left-hand side of this formula.}]{AVBW}.

We cannot rush to conclude that $I=\hat I$\,: the former contour deformation
is ill-justified for the divergent integral $I$ itself.
On the other hand, we find that it simplifies future steps
to use the transformation formula (\ref{AVT1}) which turns (\ref{AVTP}) 
into the expression (\ref{AVTR}), \emph{but still for} $\hat I$.

Next, we continue (\ref{AVTR}) to the region $\{ v \le 2 \sqrt \l \}$
($k$ pure-imaginary) by means of the transformation (\ref{AVT2}), 
which results in the expression (\ref{AVTI}) \emph{again for} $\hat I$.
Only then are we able to probe the $\l \to +\infty$ behavior of~$\hat I$
at fixed~$v \,$: using 
$\tilde k = \frac{1}{\sqrt 2} - \frac{1}{4\sqrt 2} \, v \l ^{-1/2} + O(\l ^{-1})$
and (\ref{AVE2})--(\ref{AVED}), we obtain
\bea
\hat I(\l ) \si \sim\se \textstyle \frac{2}{3} \l ^{3/4} 
\bigl[ K(\sq2) + \frac{\textstyle \d K}{\textstyle \d k} (\sq2) \, 
\frac{-1}{4\sqrt 2} \, v \l ^{-1/2} + O(\l ^{-1}) \bigr] \nonumber\\ *
&& \textstyle + \frac{1}{3} \, v \l ^{1/4} \bigl[ K(\sq2) + O(\l ^{-1/2}) \bigr]
- \frac{2}{3} \, v \l ^{1/4} \bigl[ E(\sq2) + O(\l ^{-1/2}) \bigr] \nonumber\\
\si \sim\se {\textstyle \frac{2}{3}} \, K(\sq2) \, \l ^{3/4} 
- \frac{\pi}{4 K(\sq2) }\, v \l ^{1/4} + O(\l ^{-1/4}) \qquad (\l \to +\infty );
\eea
it has no constant ($\propto \l ^0$) term, hence indeed $\hat I \equiv I$, 
the wanted canonical primitive as defined initially, cf.~(\ref{AVDC}). QED.

\noindent \emph{Remark 1.} 
The trivial outcome ($I=\hat I$) \emph{seems} to justify 
the above contour deformation directly for the divergent integral (\ref{AVTQ}), 
but this is misleading: 
our $\Pi(q)$ kept a \emph{null residue} $\beta_{-1}(0)$,
like all even $\Pi(q)$ with ${N \equiv 0 \bmod 4 \,}$; 
but generically, $\beta_{-1}(0) \ne 0$
(e.g., already for trinomial even $\Pi(q)^2$ but with ${N=6, \, 10, \ldots}$),
and nontrivial integration constants $I-\hat I \ne 0$ ought to follow.

\subsubsection*{Applications}

We can first verify (\ref{AVTQ})--(\ref{AVTI}) upon special cases, known earlier:
\smallskip

\begin{tabular}{lll}
- $\l =0:$ & $I= - \frac{1}{3} \, v^{3/2} E(1) = -\frac{1}{3} \, v^{3/2}$ 
& by~(\ref{AVEA}) for $E(1)$, cf.~(\ref{AVB4});
\\[3pt]

- $v=0:$ & $I = \frac{2}{3} \, K(\sq2) \, \l ^{3/4} 
= \frac{\G (1/4)^2}{6 \sqrt \pi} \, \l ^{3/4}$
& by~(\ref{AVE2}), cf.~(\ref{AVBH}) for $N=4 \, ;$ \\[3pt]
- $ v=2 \sqrt \l :$ & $I = \frac{\sqrt 2}{3} \, v^{3/2} [K(0)-E(0)] \equiv 0$
& by~(\ref{AVE0}), cf.~(\ref{AVTT}).
\end{tabular}
\smallskip

But above all, we can use the exact expression (\ref{AVTR}) 
to check the $v \to +\infty$ behavior of $I$ directly. 
Earlier, we predicted the asymptotic form for the general trinomial case to be, 
for $v \to +\infty$ at fixed $\l $, \cite[equation~(4.16)]{AVK}
\bea
\label{AVTA}
\int_0^{+\infty} \! (q^N+vq^M+\lambda)^{1/2} \, \d q 
\si \sim\se \! \int_0^{+\infty} \! (q^N+vq^M)^{1/2} \,\d q \, +
\!\int_0^{+\infty} \! (vq^M+\lambda)^{1/2} \,\d q \ \qquad \\
&& \qquad {}+ \delta_{M,2} \, \frac{N}{4(N \!-\! 2)} \, v^{-1/2} \lambda 
(\log v + 2 \log 2) , \nonumber
\eea
where the first line is to be made explicit through (\ref{AVBN})--(\ref{AVB4}),
and $\delta$ (last line) is the Kronecker delta symbol.

However, our derivation of (\ref{AVTA}) was quite indirect, 
and lacked independent tests. 
Now the present results allow such a test: in the quartic case,
we can directly expand~$I$ in its exact form (\ref{AVTR}) for $v \to +\infty$, 
i.e., $k \to 1^-$, and
\beq
\label{AVTK}
k' \equiv \textstyle 2 \bigl( \sqrt \l / v \bigl) ^{1/2} 
\bigl( 1+2 \sqrt \l / v \bigr) ^{-1/2} \to 0^+ .
\eeq
Then, using (\ref{AVEA}), $I \equiv -\frac{4}{3} \, \l ^{3/4} \, k'^{-3} \, 
\bigl[ (2-k'^2) \, E(k) - k'^2 K(k) \bigr]$ expands as 
\beq
I \sim \textstyle 
-\frac{8}{3} \, \l ^{3/4} \, k'^{-3} \bigl[ 1 - \frac{3}{4} \, k'^2 
- \frac{3}{16} \bigl( \log \frac{4}{k'} - \frac{1}{4} \bigr) \, k'^4
+ O(k'^6 \log k') \bigr] \, ;
\eeq
the substitution of $k'$ by (\ref{AVTK}) yields the desired $v \to +\infty$ expansion 
in terms of $v^{3/2 \, -n}$ and $v^{-1/2 \, -n} \log v$, $n \in \mathbb N$
(no $v^{1/2} \log v$ term!).
Remarkably, the next subleading term (of order $v^{1/2}$) also cancels, 
so that finally
\beq
\label{AVVA}
I \sim \textstyle - \frac{1}{3} \, v^{3/2} 
- \frac{1}{4} \l \, v^{-1/2} \bigl( \log (\l /v^2) - 4\log 2 -1 \bigr) 
\ [ {}+O(v^{-3/2} \log v) ] .
\eeq
This asymptotic equivalent then \emph{identically reproduces} 
the prediction made by (\ref{AVTA}) 
for $N=4$ and $M=2$ with the help of (\ref{AVB2})--(\ref{AVB4}),
which confirms our basic earlier result \cite[equation~(4.16)]{AVK}.

\section{The $v \to \infty$ behavior of the determinants}
\label{AV3}

We return to the spectral determinants of the quantum problem (\ref{AVSE}):
\beq
\label{AVDT}
D_N^\pm(\l ,v) = {\det}^\pm( -\d^2/\d q^2 + q^N + vq^M + \l ) ,
\eeq
which are entire functions of $(\l ,v) \in {\mathbb C}^2$ \cite{AVSb}.

\subsection{Review of the $v \to +\infty$ results}
\label{AV31}

Our key intermediate result in \cite[equations~(3.10--12)]{AVK} was,
for $v \to +\infty :$
\beq
\label{AVCM}
D_N^\pm(\l ,v)\sim \e^{\int_0^{+\infty} (q^N+vq^M+\lambda)^{1/2} \d q} 
\e^{-\int_0^{+\infty} (vq^M+\lambda)^{1/2} \d q} 
{\det}^\pm ( -\d^2/\d q^2 + vq^M + \l ) .
\eeq
Now the asymptotic formula (\ref{AVTA}) reduces this to 
\beq
D_N^\pm(\l ,v) \sim 
\e^{I(v)} \e^{ \delta_{M,2} A_0(\l ,v)} {\det}^\pm( -\d^2/\d q^2 + vq^M + \l ) ,
\eeq
\beq
\textstyle I(v) = \int_0^{+\infty} (q^N+vq^M)^{1/2} \, \d q, \quad 
A_0(\l ,v) = \frac{N}{4(N-2)} (\log v + 2 \log 2) \, v^{-1/2} \l .
\eeq
On the other hand, the exact scaling laws (\ref{AVSS}) and (\ref{AVSL}) 
for $u=0$, plus
\beq
\label{AVBM}
\beta_{-1}(0) \equiv \delta_{M,2} \, \Lambda/2 \quad 
\mbox{for } \Pi(q)=(q^M+\Lambda)^{1/2} ,
\eeq
entail 
(writing $D_M^\pm(\Lambda) \equiv {\det}^\pm( -\d^2/\d q^2 + q^M + \Lambda)$):
\beq
{\det}^\pm( -\d^2/\d q^2 + vq^M + \l ) \equiv 
v^{\pm 1/[2(M+2)]} v^{-\delta_{M,2} \, v^{-1/2} \l /8} D_M^\pm(v^{-2/(M+2)} \l ) ;
\eeq
our net asymptotic result was thus \cite[equations (5.1--4)]{AVK}
\beq
\label{AVDA}
D_N^\pm(\l ,v) \sim 
\e^{I(v)} \e^{\delta_{M,2} A(\l ,v)} v^{\pm 1/[2(M+2)]} D_M^\pm(\Lambda) 
\qquad (v \to +\infty) ,
\eeq
with $I(v) = \int_0^{+\infty} (q^N+vq^M)^{1/2} \, \d q$ given 
by (\ref{AVBN}) if $j=\frac{N+2}{2(N-M)} \notin \mathbb N$, 
or by (\ref{AVBA}) otherwise, and
\bea
\label{AVD2}
\delta_{M,2} \, A(\l ,v) \si =\se \delta_{M,2} \, 
\textstyle \frac{1}{8(N-2)} \, [(N+2) \log v + 4N \log 2] \, \Lambda , \\
\label{AVLB}
\Lambda \si \defi\se v^{-2/(M+2)} \l \qquad (\equiv v^{-1/2} \l 
\mbox{ in (\ref{AVD2}), used when } M=2) . \quad
\eea

\subsection{Extension to a sector in the complex $v$-plane}
\label{AV32}

The key to our proof of the asymptotic formula (\ref{AVCM}) for positive $v \to +\infty$ 
was \cite[\S~3.2]{AVK}
that a solution $\Psi_\lambda (q,v)$ of (\ref{AVSE}) with a recessive WKB form for $q \to +\infty$
\emph{connects} all the way down (in that WKB form) to a region $\{ 1 \ll q \ll v^{1/(N-M)} \}$
-- where it then tends to a similarly recessive solution $\Psi_{0,\l } (q,v)$ 
of the \emph{uncoupled} Schr\"odinger equation 
$[ -(\d^2 / \d q^2)+ v q^M  + \l ] \Psi_{0,\l }(q)=0$.

\begin{figure}[h]
\includegraphics[width=\textwidth]{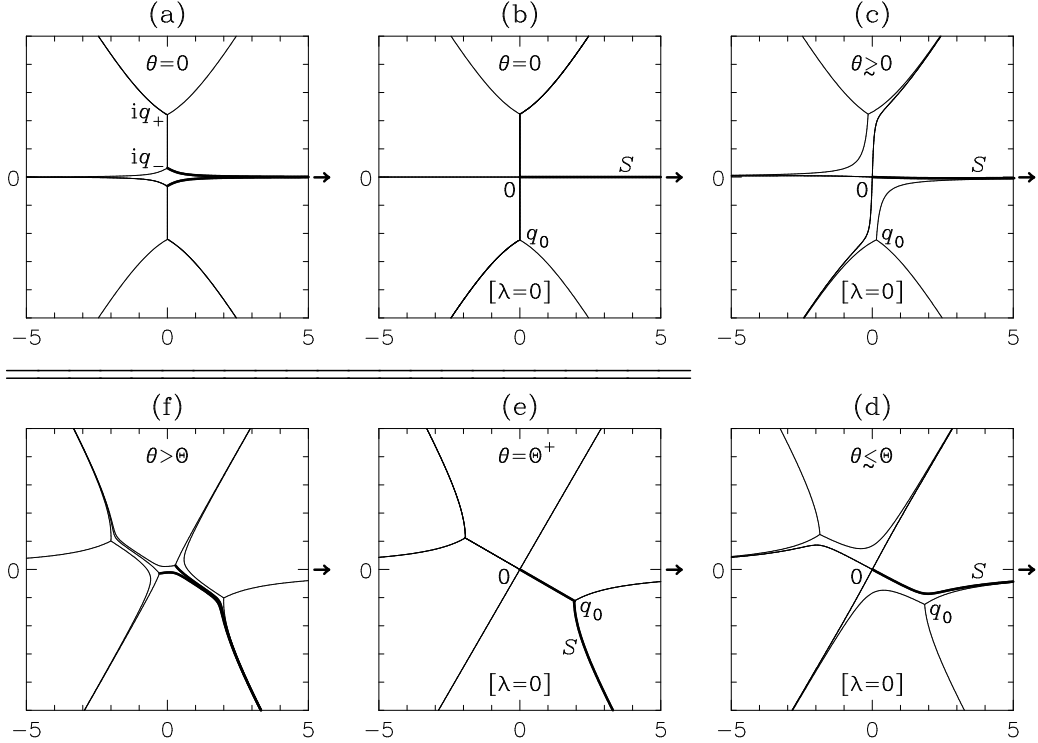}
\caption{Plots of the Stokes geometry in the complex $q$-plane for 
$\Pi(q)^2 = q^4+vq^2+\l $ and large complex $v$, 
ordered \emph{clockwise} with increasing $\theta=\arg v$ ($|v|=5$, $\l =0.5$). 
The intermediate plots (b--e) set $\l =0$ to emulate the $|v|=\infty$ regime
at finite $q \,$;
in that limit the Stokes curve $S$ (\emph{bold line}) stays linked to $q=+\infty$ 
(\emph{arrow}) for $\theta<\Theta$ (here $\Theta=2\pi/3$, by (\ref{AVCS})).
}
\end{figure}

In the complex domain, a simple sufficient condition for the WKB form to be preserved 
is for $q \in \mathbb C$ to stay within \emph{one Stokes region} 
of the momentum function $\Pi(q)$ \cite{AVQO}. 
In terms of $\theta \defi \arg v$, the above connection condition then becomes 
that the Stokes \emph{region} containing 
$\{ 1 \ll \e^{ \mi \theta /(M+2)} q \ll v^{1/(N-M)} \}$ 
(rotation given by the uncoupled equation) should \emph{link to} ${q=+\infty}$. 
E.g., when $v>0$ the \emph{central} Stokes region does include all of~$\mathbb R$, cf. Fig.~1(a).
We now need to describe the Stokes geometry for $\Pi(q)^2 = q^N + vq^M + \l $
with \emph{complex} $v \to \infty \,$; Fig.~1 illustrates the case $N=4$, $M=2$.

When $|v| \to \infty \,$: the approximate factorization of $\Pi(q)^2$ 
as ${(q^{N-M}+v)} \times {(q^M+\l /v)}$ 
makes $M$ of its complex turning points $q_j$ shrink ($\asymp v^{-1/M}$, ``inner" roots)
and the other $(N-M)$ grow ($\asymp v^{1/(N-M)}$, ``outer" roots); 
moreover, the central Stokes region contracts to a symmetrical pair 
of Stokes \emph{curves} from $q=0$ 
for the \emph{zero-energy} momentum $\Pi_{\l =0}(q)^2 = q^N + vq^M$, 
and we are to follow the ($\theta$-dependent) Stokes curve $S$ 
which starts as $S={\mathbb R}^+$ when $\theta =0$, cf. Fig.~1(b).
In the large-$v$ limit, the connection condition is that
 $\theta $ can be increased above~0 as long as 
$S$ \emph{remains linked to} $q=+\infty$ (Fig.~1(c--d);
the complex-conjugate picture results for $\theta <0$).

Following \cite[\S~3]{AVDD}, the connection condition breaks (cf. Fig.~1(e)) 
when the action integral ${\mathcal I} = \int_0^{q_0} (q^N + vq^M)^{1/2} \, \d q$ 
becomes \emph{real}, where $q_0$ is the first outer turning point met by $S$ 
as $\theta$ recedes from~0. 
That action, of \emph{instanton} type \cite{AVZJ}, is computable in closed form: 
$q_0=\e^{- \mi \, \pi /(N-M)} v^{1/(N-M)}$, 
and ${\mathcal I} = \frac{\sqrt \pi}{N+2} \,
\Gamma ( \frac{M+2}{2(N-M)} ) / \Gamma ( \frac{N+2}{2(N-M)} )
\ [\e^{-\mi \, (M+2) \pi} v^{N+2}]^{1/[2(N-M)]}$, which turns real first at
$\arg v = \frac{M+2}{N+2} \, \pi$. 
Consequently, all of \S~\ref{AV31} extends to the asymptotic sector
\beq
\label{AVCS}
{\Sigma} = \{ v \to \infty, \ | \arg v | < \Theta \}, \qquad 
\Theta = \frac{M+2}{N+2} \, \pi .
\eeq

\noindent \emph{Remark 2.}
Some examples (with $\l \equiv 0$) make us hope that our end asymptotic formula 
(\ref{AVDA}) might actually hold up to $|\arg v| < \pi$. 
1)~For ${\rm Qi}^\pm(v) \defi {\det}^\pm (-\d^2 / \d q^2 + q^4 + v q^2)$, 
this was suggested by our numerical observations \cite[equation~(87) vs Fig.~1]{AVQ} 
that ${\rm Qi}^\pm(v)$ behave analogously to the Airy functions ${\rm Ai}(v)$, ${\rm Ai}'(v)$ 
for $v \to -\infty$ as well (even though $\Theta = 2\pi/3$ only).\break
2)~The \emph{supersymmetric} determinants ${\det}^\pm (-\d^2/\d q^2 + q^N + vq^{N/2 \, -1})$ 
are known in closed form, essentially as inverse $\Gamma$-functions of $v$ 
\cite[equation~(120)]{AVQ}: their large-$v$ asymptotics then amount to the Stirling formula, 
and the latter definitely holds for $|\arg v| < \pi$ (vs $\Theta = \pi/2$).

\section{Asymptotics and the functional relation}
\label{AV4}

An early puzzle of general exact quantization conditions was 
their \emph{breakdown} (both analytical and numerical) for potentials 
$q^N+vq^2$ in the regime $v \gg 1$ (as seen for $N=4$ \cite{AVEX}). 
Naively, convergence to the elementary harmonic ($vq^2$) behavior 
would have been expected. 
We can now show that singularity to be \emph{unessential}:
i.e., the original functional relations (\emph{Wronskian identities}) 
which produce those quantization conditions behave as well as possible when ${v \to +\infty}$
for any potential $q^N+vq^M$, \emph{currently under the restriction} ${2M+2>N}$ 
(which encompasses $q^4+vq^2$, for instance).

\subsection{The basic Wronskian identity}

The spectral determinants for a general polynomial potential $V(q)$ of degree~$N$ 
obey the bilinear functional relation: \cite[equation~(40)]{AVEX}
\beq
\label{AVW}
\e^{+\mi \, \varphi_N /4} D^{+[1]} D^- - \e^{-\mi \, \varphi_N/4} D^+ D^{-[1]} 
\equiv 2 \, \mi \, \e^{\mi \, \varphi_N \beta_{-1}(0) /2} \, ,
\eeq
where $D^{\pm [1]}$ are the determinants for the \emph{first conjugate problem}:
\cite[\S~7]{AVSb}
\beq
\label{AVS}
V(q) \mapsto V^{[1]}(q) \defi \e^{-\mi \, \varphi_N} V(\e^{-\mi \, \varphi_N/2}q)
\quad \mbox{and} \quad \l \mapsto \l ^{[1]} \defi \e^{-\mi \, \varphi_N} \l ,
\eeq
\beq
\mbox{with} \quad \varphi_N \defi \frac{4 \pi}{N+2}: 
\qquad \mbox{the \emph{symmetry angle} in degree } N .
\eeq
Equation (\ref{AVW}) is but a \emph{Wronskian identity} 
for the Schr\"odinger equation (\ref{AVSE}), 
yet it has a key \emph{dynamical} role: while it seems underdetermined,
it implies a \emph{complete} set of \emph{exact quantization conditions},
which then solve (\ref{AVSE}) exactly.

A certain iterate of the transformation (\ref{AVS}) is the identity,
hence (\ref{AVW}) has a cyclic symmetry group,
specifically of order $(\hf N+1)$ when $V$ is even.
\smallskip

For the trinomial determinants (\ref{AVDT}), the first-conjugate parameters are
\beq
\label{AVV}
\l ^{[1]} = \e^{-\mi \, \varphi_N} \l , \quad v^{[1]} = \e^{\mi \, \pi /j} v , 
\quad \mbox{with } j \equiv \frac{N+2}{2(N-M)} \quad \mbox{as in (\ref{AVBN})}.
\eeq

\subsection{The $v \to \infty$ transition}
\label{AV42}

According to (\ref{AVDA})--(\ref{AVLB}) with $\Lambda = v^{-2/(M+2)} \l \,$, 
$D_N^\pm (\l ,v)$ for finite $v$ is a deformation from $D_M^\pm (\Lambda)$ at $v=+\infty$,
but the key parameter in the dynamical functional relation (\ref{AVW}),
namely the degree of $V$, and often the residue $\beta_{-1}(0)$ as well
\cite[\S~3.1]{AVK}, suffer sharp jumps at $v=\infty$. 
It is then a non-trivial task to find out whether the basic identity (\ref{AVW}) 
for $D_N^\pm$ \emph{continuously evolves} into its counterpart for $D_M^\pm$ 
in the $v \to +\infty$ limit of~(\ref{AVDA}), or not.
\smallskip

Under ${\textstyle \l \choose \textstyle v} \mapsto 
{\textstyle \l^{[1]} \choose \textstyle v^{[1]}}$ as in (\ref{AVV}),
the rescaled spectral parameter $\Lambda$ maps~to
\beq
\label{AVRM}
\textstyle \Lambda \mapsto \Lambda^{[1]} =
\exp \bigl[ -\frac{2}{M+2} \, \frac{N-M}{2} \, \mi \, \varphi_N 
- \mi \, \varphi_N \bigr] \ \Lambda = \e^{-\mi \, \varphi_M} \! \Lambda \, ;
\eeq
already this is the correct rotation angle for the limiting determinants $D_M^\pm$.

To get the asymptotic form of (\ref{AVW}) with $D^\pm \equiv D_N^\pm(\l ,v)$,
we let $v \to \infty$ in its left-hand side with $\arg v = -\pi/ \, 2j$, 
$\arg v^{[1]} = +\pi/ \, 2j$, and we invoke~(\ref{AVDA}).
\emph{The latter, by (\ref{AVCS}), requires} $\pi/ \, 2j < \Theta \ \Leftrightarrow \ j>1$ or $2M+2>N$
(otherwise the calculation will still work, but only formally until (\ref{AVCS})
extends to a wider sector). The left-hand side of (\ref{AVW}) thus displays the asymptotic form
\bea
\label{AVWA}
&& \exp \, [I(v)+I(v^{[1]})] 
\ \exp \, \delta_{M,2} [A(\Lambda ,v) + A(\Lambda^{[1]} ,v^{[1]})] \times 
\nonumber\\
&& \qquad \qquad \qquad \qquad \qquad [ z \, D_M^+(\Lambda^{[1]}) D_M^-(\Lambda)
- z^{-1} D_M^+(\Lambda) D_M^-(\Lambda^{[1]}) ] , \qquad \\
&& z \defi \e^{+\mi \, \varphi_N /4} \, \bigl( v^{[1]} / v \bigr)^{+1/[2(M+2)]} 
\qquad \mbox{(a pure phase)}. \nonumber
\eea

We now evaluate all the terms in~(\ref{AVWA}): first,
\[
\begin{array}{rcll}
I(v) \si \propto\se v^j \qquad [\, \Rightarrow \quad I(v^{[1]}) = -I(v)] 
& \mbox{if } j \notin {\mathbb N}
\mbox{ (cf.~(\ref{AVBN}), with } \beta_{-1}(0) = 0) \\[6pt]
\si =\se -\frac{2j}{N+2} \, \beta_{-1}(0) \, [\log v + {\rm const.}]
& \mbox{if } j \in {\mathbb N}
\mbox{ (cf.~(\ref{AVBA}), with } \beta_{-1}(0) \propto v^j) \\[2pt]
&&& \Rightarrow  I(v)+I(v^{[1]}) = 
\frac{2j}{N+2} \, \beta_{-1}(0) \, \mi \, \frac{\pi}{j} 
\end{array}
\]
\beq
\Rightarrow \qquad I(v)+I(v^{[1]}) \equiv \mi \, \varphi_N \beta_{-1}(0) /2 \quad 
\mbox{in all cases, cf.~(\ref{AVV})}.
\eeq
Next, just as for (\ref{AVRM}),
\beq
z = \e^{\mi \, \varphi_N /4} \e^{\mi \pi /[2(M+2)j]} =
\e^{\mi \, \varphi_N (N+2) /[4(M+2)]} \equiv \e^{\mi \, \varphi_M /4} .
\eeq
Finally, and only relevant when $M=2$, in which case $\Lambda^{[1]}=-\Lambda$,
\beq
\label{AVHA}
A(\Lambda ,v) + A(\Lambda^{[1]} ,v^{[1]}) 
= \textstyle -\frac{N+2}{8(N-2)} \, \mi \, \frac{\pi}{j} \, \Lambda 
= -\frac{N+2}{16} \, \mi \, \varphi_N \, \Lambda \equiv -\mi \, \pi \Lambda /4 .
\eeq

In the end, substituting (\ref{AVRM})--(\ref{AVHA}) into (\ref{AVW}) we indeed get
\beq
\label{AVH}
\e^{+\mi \, \varphi_M /4} D_M^+(\e^{-\mi \, \varphi_M} \! \Lambda) D_M^-(\Lambda) 
- \e^{-\mi \, \varphi_M/4} D_M^+(\Lambda) D_M^-(\e^{-\mi \, \varphi_M} \! \Lambda)
\equiv 2 \, \mi \e^{\delta_{M,2} \, \mi \, \pi \Lambda /4} ,
\eeq
which \emph{is the correct form} of (\ref{AVW}) for 
$D_M^\pm(\Lambda) = {\det}^\pm( -\d^2 /\d q^2 +q^M+\Lambda)$
(whose $\beta_{-1}(0)$ is given by (\ref{AVBM})). For more details:
if $M>2$, see \cite[equation~(5.32)]{AVZ}; if ${M=2}$, then
\beq
\textstyle
D_2^+(\Lambda) = 2^{1-\Lambda/2} \sqrt \pi /\G (\frac{1+\Lambda}{4}), \qquad
D_2^-(\Lambda) = 2^{-\Lambda/2} \sqrt \pi /\G (\frac{3+\Lambda}{4}),
\eeq
and (\ref{AVH}) with its ``anomalous" right-hand side boils down to 
the reflection formula for the Gamma function; 
the harmonic-oscillator quantization condition can then also be recovered 
solely from (\ref{AVH}) \cite[Appendix~A.2.3]{AVQ}.
\smallskip

In conclusion, we have verified that the exact functional relation (\ref{AVW}), 
governing both $D_N^\pm(\l ,v)$ and $D_M^\pm(\Lambda)$ (cf.~(\ref{AVH})), 
is compatible with the general perturbation formula (\ref{AVDA}),
\emph{currently under the restriction} $2M+2> N$, or $j>1$
(which includes the quartic oscillators):
this further validates the exact-WKB description of perturbative regimes in~\cite{AVK}. 
Remaining desirable tasks are: 1) to lift the restriction $j>1$ 
(e.g., by extending (\ref{AVDA}) to $\{ |\arg v|<\pi \}$,
cf. Remark~2 in \S~\ref{AV32}); 
and 2) to find exact quantization conditions that themselves behave \emph{continuously}
in the zero-coupling limit (here, $v=+\infty$).

\end{document}